\documentstyle[aaspp4,tighten]{article}
\input{psfig}
\def\apj #1 {{ApJ} { #1}, }
\def\apjsupp #1 {{ApJS} { #1}, }
\def\apjl #1 {{ApJ}  { #1}, }
\def\aj  #1 {{AJ} { #1}, }
\def\aa  #1 {{A\&A} { #1}, }
\def\aarev  #1 {{A\&AR} { #1}, }
\def\aasupp  #1 {{A\&AS} { #1}, }
\def\mn  #1 {{MNRAS} { #1}, }
\def\pasp  #1 {{PASP} { #1}, }
\def\paspc  #1 {{PASPC} { #1}, }
\def\pasj  #1 {{PASJ} { #1}, }
\def\revmex  #1 {{RevMexAA} { #1}, }
\def\jaa #1 {{JA\&A} { #1}, }
\def\annrev #1 {{ARA\&A} { #1}, }
\def\apss #1 {{Ap\&SS} { #1}, }
\def\baas #1 {{bAAS} { #1}, }
\def\nat #1 {{Nat} { #1}, }
\def\obs #1 {{Observatory} { #1}, }
\def\memital #1 {{Mem. Soc. Astron. Ital.} { #1}, }
\def\msol {\rm M$_\odot$}                              

\def\intens {10$^{-7}$ erg cm$^{-2}$ s$^{-1}$ sr$^{-1}$}
\def\menmas #1 #2 {{$_{-#1} ^{+#2}$}}

\def\col {\hfil & \hfil}
\def\raya {\noalign {\medskip \hrule \medskip}}

\newcommand{\beq}{\begin{equation}}
\newcommand{\eeq}{\end{equation}}

\begin{document}

\def\etal{{\it et~al.\ }}
\def\eg{{\it e.~g.\ }}
\def\ie{{\it i.~e.,\ }}

\title{The warm interstellar medium around the Cygnus Loop$^*$}

\author{Joaqu{\'\i}n Bohigas\altaffilmark{1,3}, 
Jean Luc Sauvageot\altaffilmark{2,4} and
Anne Decourchelle\altaffilmark{2,5}}

\altaffiltext{1}{Instituto de Astronom{\'\i}a, UNAM, Apdo. Postal 877, 
22830 Ensenada, B.C., M\'exico}
\altaffiltext{2}{C.E.A., DSM, DAPNIA, Service d'Astrophysique, C.E. 
Saclay, F91191, Gif sur Yvette Cedex, France}
\altaffiltext{3}{Email address: jbb@bufadora.astrosen.unam.mx}
\altaffiltext{4}{Email address: jsauvageot@cea.fr}
\altaffiltext{5}{Email address: adecourchelle@cea.fr}


$^*$Based on observations collected at the Observatorio Astron\'omico
Nacional, San Pedro M\'artir, B.C., Mexico.

\begin{abstract}
\noindent

Observations of the oxygen lines [OII]3729 and [OIII]5007 in
the medium immediately beyond the Cygnus Loop supernova remnant
were carried out with the scanning Fabry-P\'erot
spectrophotometer ESOP. Both lines were detected in three
different directions - east, northeast and southwest - and
up to a distance of 15 pc from the shock front. The ionized medium 
is in the immediate vicinity of the remnant, as evinced by the 
smooth brightening of both lines as the adiabatic shock transition 
(defined by the X-ray perimeter) is crossed. These lines
are usually brighter around the Cygnus Loop than in the general 
background in directions where the galactic latitude is 
$\geq ~ 5^ \circ$.  There is also marginal (but significant) 
evidence that the degree of ionization is somewhat larger 
around the Cygnus Loop.  We conclude that the 
energy necessary to ionize this large bubble of gas could have 
been supplied by an O8 or O9 type progenitor or the particles heated
by the expanding shock front. 
The second possibility, though
highly atractive, would have to be assessed by extensive modelling.

\end{abstract}


\section{Introduction} 

An understanding of the properties and physics of the medium
where supernova remnants (SNR's) expand is essential in order to
develop consistent scenarios for their evolution and physical 
structure. In turn, the sequence of events leading to a supernova 
remnant and its demise define to a very large extent the structure, 
physical and chemical state and evolution of the interstellar medium.
The Cygnus Loop is particularly well suited to study these
questions. Being in an advanced evolutionary stage, different
regions of the remnant are found interacting with different components
of the interstellar medium. The remnant is close enough to study 
in great detail these interactions. In some circumstances the 
structure of the surrounding medium is deduced from observations
of the SNR in different spectral domains (e.g. Graham {\it et al.} 1991; 
Hester, Raymond \& Blair 1994; Decourchelle {\it et al.} 1997). When 
dealing with atomic or molecular gas this information has been 
gathered directly (e.g. DeNoyer 1975; Scoville {\it et al.} 1977). 
Direct observations of ionized gas in the surrounding medium are 
considerably more complicated, since the emission lines arising from 
these regions are bound to be faint and prone to background confusion.
In this respect the Cygnus Loop offers a substantial advantage, since
it is placed at a large galactic latitude ({\it b} = -8.6$^\circ$).

Observations of faint emission lines in diffuse media have been
carried out with a scanning Fabry-P\'erot spectrophotometer. This 
technique has been successfully applied when searching for emission from 
Fe$^{+9}$ and Fe$^{+13}$ in SNR's (Ballet {\it et al.} 1989; Sauvageot 
{\it et al.} 1990; Sauvageot \& Decouchelle 1995), 
or exploring line emission in the warm 
component of the interstellar medium (e.g. Reynolds 1983, 1985). In this 
paper we searched for line emission from warm ionized gas 
in the direction of the Cygnus Loop using such an instrument, 
ESOP (Dubreuil {\it et al.} 1995). In comparison with 
a grating instrument, ESOP can sample a large solid angle with a 
substantial luminosity advantage, at the cost of a reduced spectral 
range. It has no spatial resolution and is less efficient than 
direct imaging observations, but the line of interest can be 
isolated from other spectral features and the underlying continuum. 
This is a precious advantage when the line is faint. 

A description of the instrumental setup and the process followed in
data reduction is presented in $\S$2. Results are described in detail 
in $\S$3, and a discussion on the viability of several possible 
ionizing sources is given in $\S$4. Finally, conclusions and research
perspectives are summarized in $\S$5.

\section{Observations and data reduction}

ESOP was mounted on the 1.5 meter telescope of the Observatorio 
Astron\'omico Nacional at San Pedro M\'artir, B.C., M\'exico. Data
was gathered in four runs: July 1994, October 1994, July 1995 and
August 1996. 
The observations reported in this paper are for the [OII] lines 
at 3726 and 3729 \AA, [OIII] at 5007 \AA~and HeI 5876 \AA.
We used narrow band ($\sim$ 15 \AA~FWHM) interference filters centered at 
these lines when T = 20.5 $^\circ$C. Each observation consists of 50 to 100 
scans of a hundred steps each. The integration time at each step is 170 ms.
Thus, the typical total integration time for each data point is around
20 minutes.
The free spectral range (FSR) was 11.3 \AA~during the first [OII] 
observations and 14.5 \AA~in the following ones, in both cases
centered at 3727 \AA~(with the [O II] filter's FWHM = 15 \AA).
The FSR was not completely scanned by the Fabry-P\'erot in the
first [OII] observations. All [OIII] observations were conducted with 
a FSR of 10.8 \AA~centered at 5007 \AA~(with the [O III] filter's 
FWHM = 15 \AA). Because our filters are well centered on the oxygen
lines and the FSR is greater than half the filters' bandwidth,
these are not contaminated. Sky lines
lying far from the center of the filter, may have a contribution
from another order.
Since temperature variations shift the filter's transmission 
curve and humidity affects the air capacitance in the FP gap, the 
instrument works in a dry nitrogen atmosphere and is under temperature 
control (at T $\simeq~20.5~^\circ$C, with $\delta$T $\simeq ~ 0.2~^\circ$C).
White and comparison lamps were regularly measured 
in order to follow any bandpass drift in the filter or any variation in the
FP gap. All observations were carried out with a 150" circular diaphragm.

Data reduction and analysis has been amply described in 
Ballet {\it et al.} (1989) and Sauvageot and Decourchelle (1995). It takes
into account that the geocoronal contribution - lines and continuum - 
is unstable. No attempt is made to subtract a blank sky directly. Instead, 
the observation is modelled with a combination of sky and source lines 
and continuum. $\chi^2$ tests are performed to estimate the statistical
error of measurements. 

Sky continuum and lines must be monitored carefully in this type of
observations. Typical spectra at the medium contiguous to the Cygnus
Loop and randomly selected positions are presented in Figures 1 
and 2 ([OII] and [OIII]). 

Though absolute wavelength calibrations are wrong, relative ones (meaning
dispersion) are correct. This is not consequential since this paper is 
only concerned with photometric results and there can hardly be
any confusion regarding the identification of the line given its
relative prominence (for instance, the bright feature at 5010.94 \AA~
in Figure 2b is obviously the [OIII]5007 line). For further clarity,
arrows located on the upper part of Figures 1a, 1b, 2a and 2b indicate
which is the line that is being studied.
As can be seen, the O$^+$ lines are easily recognized and discriminated.
Regarding the [OIII] filter, there can hardly be a confusion on the
identification and measurement of [OIII]5007 since sky lines were found 
to be very weak in this spectral region. 
Since HeI 5876 was not detected outside the bright
optical filaments of the Cygnus Loop, a discussion on the sky 
lines admitted by this filter is not necessary.
Our observations were flux calibrated observing standard stars 29 Pisc and 
58 Aql several times during the night. For the [OIII] calibration we had
to take into account the contamination from the other order to the
standard star continuum. At its worst, we estimate that
the absolute values for the specific intensity are accurate at the 25$\%$
level, comparable to our statistical error bars.

\section{Results} 

Regions within the Cygnus Loop and the medium surrounding it were observed 
in three directions (northeast, east and southwest), along lines 
approximately perpendicular to the shock front. Our observational
pointings are shown in Figure 3. The shock front is defined 
as the outer X-ray boundary of the supernova remnant (Ku {\it et al.} 1984; 
Ballet $\&$ Rothenflug 1989). Our results for the O$^+$ and O$^{+2}$ 
lines are compiled in Tables 1, 2 and 3 (NE, E and SW traces). 
The information contained in these 
tables is as follows: code name and coordinates for each position, angular 
distance ($\delta$, in arcseconds) between the observed position and the 
shock front (negative values for regions inside the SNR), angular distance 
from the center of the SNR in terms of its radius, 1 + $\delta / \theta$, 
where $\theta$ = 5850" is the angular radius of the Cygnus Loop
(mean value of the semi-axes, Green 1988), the specific intensity of
[OII]3729 \AA~and [OIII]5007 \AA~(henceforth [OII]3729 and
[OIII]5007) in \intens, and the ratio of these lines (henceforth
3729/5007). Except for pointings in the optical filaments of the Cygnus 
Loop (such as NE0, NE1, E0, E1 and SW0), the ratio [OII]3727/3729 is 
always at the low density limit ($\simeq$ 0.67, N$_e \leq$ 50 cm$^{-3}$), 
so there is no need to report the flux of the other [OII] line.
Upper and lower bounds for the specific intensity are for a 90$\%$ confidence 
level in the fitting procedure. Uncertainties on the 3729/5007 line ratio
were determined from these bounds. 
Both lines were detected at every position, some of them
very far from the shock front: for instance, region E26 is 3894" distant 
from it, $\simeq$ 15 pc at the distance where the Cygnus Loop is (770 pc). 
Except for the remnant's optical filaments, the HeI line was not detected.

The NE and E traces of [OII]3729 and [OIII]5007 as a function 
of 1 + $\delta / \theta$ are plotted in Figure 4. As can be seen, both
traces are nearly identical, and the specific intensities well away 
from the shock front are roughly constant (particularly [OII]3729). 
Notice that the adiabatic shock transition is revealed as 
[OII]3729 and [OIII]5007 brighten up smoothly as the X-ray perimeter
of the SNR is crossed (at 1 + $\delta / \theta$ = 1). 
This brightening is caused by adiabatic compression of
the plasma and enhanced line emissivities, which overcompensate 
the decreasing concentration of both ions since these are now immersed 
in a higher temperature medium. The SW trace is not
plotted, but our measurements (see Table 4) also reveal the shock transition. 
The adiabatic transition is apparently complete at 1 + $\delta / \theta 
\simeq$ 0.95 or $\simeq$ 1 pc inside the shock front. This is the
most compelling evidence that this medium is not a projected HII region or 
parcel from the general background, but is in the immediate vicinity of 
the Cygnus Loop. Our most distant pointing (region E26) is $\simeq$ 
36 pc away from the remnant's center, which implies that the 
surrounding medium is ionized up to a distance of at least $\simeq$ 50 pc.
An additional and less conspicuous feature is that the
specific intensity of both lines seems to rise slightly even before
the shock front is crossed (from 1 + $\delta / \theta~\simeq$ 1.1, or 
$\simeq$ 2 pc away from the shock). This seems to indicate that the 
plasma immediately beyond the edge of the Cygnus Loop is being affected 
by the SNR before it encounters the blast wave, though this turn up
may also be due to irregularities in the boundary of the remnant.

As mentioned above, [OII]3729 maintains an approximately constant level
beyond $\simeq$ 0.1 - 0.2 times the radius of the SNR. This level is 
nearly identical in the NE and E directions (for which we have sufficient
data points). It is worth noticing that [OII]3729 is practically the 
same in regions separated by as much as $\simeq$ 5800" (NE22 and E26), 
some 22 pc at the distance to the Cygnus Loop. Thus, the [OII]3729 data indicates that the medium beyond the shock front of the Cygnus Loop 
is very extended, and quite possibly surrounds the eastern face of the 
remnant. Furthermore, the data for the SW trace, albeit limited,
suggests that this medium exists all around the Cygnus Loop.
[OIII]5007 does not display such a regular behaviour. Along the NE trace 
it dims slightly at positions NE8, NE9 and NE15, but it brightens up 
again further away (NE18 and NE22). Along the eastern trace [OIII]5007 
behaves more regularly, weakening very noticeably in the two most distant 
pointings (E18 and E26). Thus, the [OIII]5007 observations imply that 
physical conditions in the surrounding medium are not homogeneous. If 
the temperature is uniform, there are variations in the oxygen degree of 
ionization: in the low density limit, N(O$^+$)/N(O$^{+2}$) $\simeq$ 
2.7, 1.8 or 1.5 in most positions, and up to 4.8, 3.3 or 2.9 at E18 
and E26 (for T$_e$ = 6000, 8000 and 10000 $^\circ$K). Alternatively, 
if the degree of ionization is constant, the temperature would have to 
be twice as large where [OIII]5007 is faintest. It seems difficult to 
maintain such high temperatures.

In order to assess if there is a difference between the medium 
surrounding the Cygnus Loop and the general background, observations 
were carried out on randomly selected directions located 
at {\it $\vert$b$\vert~\geq~5^\circ$}, as the Cygnus Loop is. Specific
intensities as low as $\sim 0.07 \times$ \intens~ were measured, which
gives an idea on the instrumental sensitivity. As expected, line brightness 
is not uniform in the galactic background. [OII]3729 was observed and
detected 13 times over the four observing runs.
Specific intensities comparable to some of those observed in
regions contiguous to the Cygnus Loop (0.89 and 0.64 $\times$ 
\intens ) were found in only two directions. Elsewhere they were
markedly smaller, and in the mean [OII]3729 = 0.47$\pm 0.17 \times$ \intens .
Out of 9 measurements, [OIII] was detected in 7 positions. [OIII]5007 =
0.68 $\times$ \intens~in the brightest sky background region, 
similar to what we found in some regions around the SNR, but in all
other directions the line was much fainter. In the mean,
[OIII]5007 = 0.25$\pm 0.24 \times$ \intens . In most sites we 
measured either one or the other line due to the complications involved 
in changing the instrumental setup. Both lines were measured only in two 
directions, where we found 3729/5007 = 3.13 and 7.80. This implies that
N(O$^+$)/N(O$^{+2}) \simeq$ 6.4 and 16.8 (for T$_e$ = 8000 $^\circ$K), 
a rather low level of ionization. HeI 5876 was not detected, confirming 
the faintness of this line in the general background (Reynolds 
\& Tufte 1995).

As far as we know, this is the first time that O$^+$ emission 
from the diffuse interstellar medium has been reported.
Reynolds (1985) observed [OIII]5007 in three directions: the
specific intensity was less than 0.5 $\times$ \intens~in 
one of them, 2 and 1.8 $\times$ \intens~in the other two. In
these two, which are amongst the brightest sky background regions
as defined by the H$\alpha$ intensity (Reynolds 1983), [OIII]5007 is
substantially brighter than anything we find beyond the Cygnus Loop.
We measured [OIII]5007 at the second brightest 
region ($\it l$ = 96.0$^\circ$, $\it b$ = 0.0$^\circ$)
and obtained 0.68 \menmas 0.15 0.20 $\times$ \intens, 2.6 times less
than Reynolds. The discrepancy is probably related to the
vast difference in aperture sizes (49' $\it vs.$ 2.5'), so that
a region with particularly intense emission was included in
Reynolds' diaphragm but not in ours.

Thus, there is a measurable background contribution to the 
intensity of the oxygen lines but, as can be seen from
Tables 1 to 3, emission is generally larger in the medium surrounding 
the Cygnus Loop: [OII]3729 is at least 1.5 times brighter 
around the Cygnus Loop than in the general backround, whereas [OIII]5007
is between 2 and 5 times more intense. Thus, the data supports the 
conclusion that [OII]3729 and [OIII]5007 in the medium just beyond the 
Cygnus Loop are usually brighter than in the general galactic 
background at least up to a distance of $\simeq$ 15 pc from
the shock front (about 0.6 times the radius of the remnant). There is
also marginal but significant evidence indicating that the degree of 
ionization is higher in the medium around the Cygnus Loop.

The absence of HeI 5876 emission merits a discussion given the detection
of [OIII]5007, which requires more energetic photons for this ion to exist 
(35.1 {\it vs.} 24.6 eV). We take notice that there is an antecedent: 
Reynolds (1985) found intense [OIII]5007 emission at 
$\it l$ = 194.0$^\circ$ $\it b$ = 0.0$^\circ$, but Reynolds \&
Tufte (1995) searched for HeI 5876 at this location with negative
results, implying that He$^+$/He $\leq$ 0.3.  In general, I(5876)/I(5007) = 
$\epsilon$(5876)/$\epsilon$(5007) He$^+$/O$^{+2}$, where 
$\epsilon$(5876) and $\epsilon$(5007) are the emissivities for
HeI 5876 and [OIII]5007. For cosmic abundances, and in the low density 
limit, I(5876)/I(5007) = (0.08, 0.02, 0.008) (He$^+$/He)(O/O$^{+2}$) 
for T = 6000, 8000 and 10000 K. The fraction of doubly ionized
oxygen is determined from the previously calculated
O$^+$/O$^{+2}$ ratio, assuming that there are no higher ionization
stages and that O$^0$/O$^{+2}$= 2. And since I(5007) $\simeq~0.9 \times$ 
\intens~in the medium surrounding the Cygnus Loop, it follows that
I(5876) $\simeq$ (0.42, 0.09, 0.03) (He$^+$/He) $\times$ \intens. 
Finally, the non-detection of HeI 5876 implies that I(5876) 
$\leq$ 0.05 $\times$ \intens (our detection limit). Consequently, 
HeI 5876 emission will be under our detection threshold if 
(He$^+$/He) $\leq$ (0.17, 0.78 or 1) for the aforementioned temperatures. 
Additionally, since He$^+$/He must be larger than O$^{+2}$/O, 
we conclude that the ambient temperature is larger than $\sim$ 6000 K.  

\section{Discussion}

The medium revealed by our observations has the general
properties of the partially ionized warm component of the interstellar 
medium, which is so favourable for the observability of supernova remnants 
(Kafatos {\it et al.} 1980). But it is different insofar as it is
generally brighter, and a couple of observations on sky background regions
also suggest that the degree of ionization, given by N(O$^+$)/N(O$^{+2}$), 
is larger around the Cygnus Loop. On the other hand we did not
detect HeI 5876 emission, and in this respect there is no difference
between the medium just beyond the Cygnus Loop and the general galactic 
background (Reynolds and Tufte 1995). But there are 
several reasons to expect somewhat different properties in the
medium surrounding the Cygnus Loop: the remnant has been included in the 
so called Cygnus superbubble, along with several OB associations,
the SN progenitor might have been a source of ionizing energy, the SN
itself produced a large amount of UV photons and, finally, ionizing
radiation is also generated by the expanding shock wave. We will
discuss these possible sources in the following paragraphs.

The Cygnus Loop is at the southern edge of the extremely rich and complex
region known as the Cygnus superbubble, which has been extensively
described and analysed by Bochkarev \& Sitnik (1985). The superbubble
has seven OB associations containing 
48 O type stars and nearly 70 B type stars. With the exception of
Cyg OB4 and Cyg OB7, all of them are at a distance of 1.2 kpc or
more. Cyg OB7 is at approximately the same distance as the Cygnus Loop, 
but is located some 20$^\circ$ away from it (about 280 pc). 
Cyg OB4 is also relatively near, but though it has been classified
as an OB association, it does not contain any O or B star. Thus, it is 
doubtful that these OB associations can account for the ionization
of the medium surrounding the Cygnus Loop. We also contemplated 
the possibility that this medium is an extended low surface brightness 
HII region produced by an early type star in the vicinity of the SNR. 
A visual inspection of POSS plates and a thorough search 
of O and B stars catalogs (Cruz-Gonz\'alez {\it et al.} 1974; Garmany,
Conti \& Chiosi 1982) renders no support to this possibility. The SAO 
catalog was also explored with negative results. 

The ionized medium around the Cygnus Loop may also be the relic HII 
region of the progenitor star. The progenitor should have produced 
P$_{UV}~\sim~5 \times 10^{48}$ N$_H ^2$ UV photons per second to 
create a 50 pc Str\"omgren sphere if the medium is fully ionized, 
as the presence of O$^{+2}$ seems to imply (but notice that 
Graham {\it et al.} (1991) found shocked H$_2$ in the Cygnus Loop). The
mean particle density in the medium where the remnant evolved at least
until recently is $\sim$ 0.1 - 0.2 cm$^{-3}$ (Ku {\it et al.} 1984;
Levenson {\it et al.} 1997), which signifies that the required spectral 
type of the progenitor star must have been earlier than or equal to B0. 
Furthermore, the 3729/5007 ratio ($\simeq$ 1 with no
reddening correction, about 1.5 for a 1 magnitude visual extinction),
implies that the effective temperature of the ionizing star is close to 
35000 K (Stasinska 1978), corresponding to a spectral type slightly
later than O8. Under this hypothesis, the mass of the progenitor 
would have been between 20 and 25 \msol. 
Such a star spends most of its lifetime as a blue giant and only
$\sim 1 \%$ of its existence ($\sim 10^5$ yr) as a red supergiant 
(Brunish \& Truran 1982). This is substantially less than the
recombination time ($\sim~10^5/N_H$ yr). Thus, an O8 or O9 progenitor
surrounded by a pervasive low density medium can account for our 
observations. 
 
On the other hand, based on the X-ray morphology of the SNR, 
Levenson {\it et al.} (1997) favor a scenario with a progenitor
of spectral type later than B0. According to them, the Cygnus Loop 
evolved within the $\sim$ 20 pc homogeneous low-density HII region created 
by this progenitor, and is now bursting into the relatively dense and inhomogeneous medium surrounding it, in the manner described by McKee 
{\it et al.} (1984). This would explain the existence of abundant local 
inhomogeneities on the external surface of a nearly circular remnant. 
Notice that an earlier type progenitor would create a larger homogeneous 
cavity.

The UV radiation produced by the SN explosion can also be an important
ionizing source, as was palpably revealed when narrow emission
lines appeared in the UV spectra of SN1987A $\sim$ 70 days after
the event (Fransson {\it et al.} 1989). At least some 10$^{44}$ erg
of ionizing energy was required to produce these lines (Fransson 
{\it et al.} 1989), substantially less than the $10^{46}-10^{47}$ erg that
hydrodynamical models had predicted for the ionizing burst of 
SN1987A (Shigeyama, Nomoto \& Hashimoto 1988; Woosley 1988). But values as
large as $10^{48}-10^{49}$ erg in ionizing energy have been mentioned in
the literature (Chevalier 1977; Chevalier 1990). If the mean photon energy 
is 20 eV, the largest radiation ``pulse" ionizes the surrounding medium
up to a distance of 14 N$_H ^{-1/3}$ pc at the most, where N$_H$ 
is the mean hydrogen density. The density would have to
be smaller than 0.02 cm$^{-3}$ in order to produce a 50 pc bubble of
ionized gas. Observations and models for the evolution of the Cygnus 
Loop lead to substantially larger mean densities in the surrounding medium
(e.g. Ku {\it et al.} 1984).
 
Ultraviolet photons are constantly being supplied by the expansion of
the SNR as the shock heated particles produce ionizing radiation as they 
move downstream. This has been discussed in the numerous radiative shock
wave models developed over the last 25 years (e.g. Cox 1972;
Dopita 1977; Raymond 1979; Cox \& Raymond 1985; Shull \& McKee 1979; 
Binette, Dopita \& Tuohy 1985; Sutherland,
Bicknell \& Dopita 1993). The effect of this ionizing 
radiation on the upstream gas has been considered in the context of AGN 
(Daltabuit \& Cox 1972) or, more recently, the emission line filaments 
in Centaurus A (Sutherland {\it et al.} 1993). A review on the many topics
opened to this question was written by Dopita (1995). But to the best of 
our knowledge, little attention has been directed to the effect of the 
photoionizing flux of SNR's on the galactic interstellar medium.
At this point we are specifically interested in determining the size
of the bubble of ionized gas that can result from the UV flux produced
by the shock heated particles, in order to establish if this energy
source is sufficient to create the large sphere of ionized gas that is
implied by our observations. An estimate of this quantity can 
be determined following a very simple line of arguments.
The number of upstream moving photons produced each second by a SNR 
expanding into a medium with density $N_0$, is given by

\beq
P_{UV} =  4 \pi R_0^2 N_0 V_0 \phi_{UV}
\eeq

where $\phi_{UV}$ is the number of upstream moving UV photons produced 
per shocked particle, and $R_0$ and $V_0$ are the remnant's radius and 
velocity. If this quantity is equal to the number of recombinations, then

\beq
(R_i/R_0)^3 = 74.8 V_7 \phi_{UV}/(N_0 R_{pc}) + 1 
\eeq

where $R_i$ is the radius of the ionized volume measured from
the remnant's center, $V_7$ is the shock velocity in 100 km s$^{-1}$
and $R_{pc}$ is the radius of the SNR in parsec. The latter can
be determined assuming that the evolution of the Cygnus Loop is
described by Sedov's (1959) solution. In the case of
the Cygnus Loop this assumption can be objectionable,
but is probably adequate given the scope of this discussion. In this
case,

\beq
R_{pc} = 19.4 (E_{50}/N_0)^{1/3} V_7^{-2/3}
\eeq

where $E_{50}$ is the kinetic energy deposited in the SNR in 
10$^{50}$ erg. Equations (2) and (3) lead to,

\beq
(R_i/R_0)^3 = 3.86 V_7 ^{5/3} \phi_{UV}/(E_{50} N_0^2)^{1/3} + 1 
\eeq

For a given metallicity the number of ionizing photons per particle 
only depends on the shock velocity. Shull $\&$ McKee (1979) present their 
results in a more 
amenable fashion than Binette {\it et al.} (1985) and Dopita (1995), and 
the following analytical approximation for $\phi_{UV}$ can be derived 
from their work,

\beq
\phi_{UV} \simeq 1.08 (V_7^2 - 0.58) 
\eeq

Their calculations stop at $V_7$ = 1.3,  but it is probably correct 
to extend this approximation to larger velocities (the functional 
dependence should not change, see Dopita 1995). For an 
equilibrium cooling function the shock can be radiative up to 
$V_7 ~ \simeq$ 1.5 - 2. But the plasma behind the shock front will 
be underionized with respect to collisional ionization equilibrium,
and in this condition cooling is more efficient (Sutherland and 
Dopita 1993). Thus, the shock will become radiative at somewhat higher 
velocities. 

The size of the region that can be ionized by photons produced by the 
shock heated particles, R$_i$, can now be determined from equations (3), 
(4) and (5). Results as a function of shock velocity are presented in 
Table 4 for various combinations of ($E_{50}$,$N_0$): (1,1), (3,0.2) 
and (1,0.2). The second set of parameters is representative of the
Cygnus Loop (Ku \etal 1984). As 
can be seen, a 50 pc ionized bubble can be produced even in the most 
conservative case. The size of the region of ionized gas is surprisingly
large when the standard parameters for the Cygnus Loop are considered.
Furthermore, since the evolutionary timescale of a SNR is much
shorter than the recombination time, it follows that the ionizing
radiation supplied by the remnant as it continues evolving would 
further increase the size of the ionized region. Consequently it
appears that, at least from the point of view of the energy budget,
radiative shock waves can produce a very extended environment of ionized
matter around them. A stricter analysis is no doubt necessary, but it seems  
improbable that it will lead to a qualitatively different conclusion
on the number of ionizing photons produced by the expanding SNR, 
and consequently on the size of the region that will be influenced by them. 
But if there seems to be little doubt that SNR's can ionize large
regions of the interstellar medium, it has to be shown that these
objects can do so in the course of their evolution. This
is an essential point in relation to this work.

It is worth pointing out that, in comparison to other ionizing sources, 
the ionizing energy produced by
SNR's can be of considerable importance. Integrating equation (1) with
the aforementioned hypotheses, it is easy to see that a SNR
will produce 1.2 $\times 10^{60} E_{50}$ UV photons as it slows
down from 250 to 80 km s$^{-1}$, $\sim 30 \%$ of its initial kinetic
energy if the mean photon energy is 15 eV. On the other hand, any
main sequence B0 or O type star will produce some 10$^{63}$ UV photons
during its lifetime. Considering that SN's are some 20 times more 
abundant than O type stars, this implies that, during their radiative
phase, SNR's will generate about a tenth of the UV flux produced by all B0 
and O type stars. This is not a small number. Furthermore, SNR's will
be a major source of UV radiation in stellar systems lacking massive stars.

\section{Conclusions}

Evidence  was presented for the existence of an extended ionized medium 
surrounding the eastern face of the Cygnus Loop, and quite possibly 
the entire remnant. The shock transition is revealed by the 
slow rise in the specific intensity of [OII]3729 and [OIII]5007 
as the X-ray perimeter of the SNR is crossed. This is indisputable 
proof that this medium is in the immediate vicinity of the Cygnus Loop. 
Our most distant pointing (region E26) is $\simeq$ 36 pc away from the 
remnant's center, which implies that there is ionized gas at least
up to a distance of $\simeq$ 50 pc. It would be interesting to observe
more distant regions, preferably in the O$^{+2}$ line, in order to see
whether there is an outer boundary or the medium merges smoothly with 
the general background. The medium around the Cygnus 
Loop is somewhat different to the general galactic backround: [OII]3729 
and [OIII]5007 are usually brighter, and there are indications 
that the degree of ionization, given by N(O$^+$)/N(O$^{+2}$), is also
larger around the SNR. On the other hand it is similar insofar as
HeI 5876 emission is also conspicuously absent.

We explored several possible sources which may produce the ionizing
energy required to account for the existence of this medium. Viable
external sources, such as an isolated early type star or an OB 
association, could not be found. We also concluded that the ionizing
radiation produced by the SN explosion was probably insufficient.
An early type (between O8 and O9, but closer to the former) progenitor 
embedded in a low density medium can account for the required energy 
budget, but  a later type progenitor has been suggested 
by Levenson {\it et al.} (1997). Finally, we
showed that the UV radiation produced by the shock heated particles 
{\it can} generate a large bubble of ionized gas, but detailed modelling 
is required in order to see if it {\it will} do so during the SNR's lifetime.

From the observational point of view, it is advisable to inspect other 
emission lines in order to explore the spectral properties of the medium 
surrounding the Cygnus Loop, and decide if it is indeed distinct
to the warm component of the interstellar medium.
Unfortunately, our instrumental resolution is insufficient to discriminate 
coronal and galactic H$\alpha$ emission, the key line in Reynolds' research 
on the properties of this component of the interstellar medium. But 
other spectral lines, such as [NII]6584 \AA~and [SII]6717,6731 \AA, 
are open to inspection since they are less affected by geocoronal emission.
Needless to say, similar observations of the medium surrounding other 
SNR's should furnish valuable information. Targets located away from the
galactic plane are preferable, since background confusion is avoided. 
Further research along
these lines may be helpful regarding the still open question on the origin 
of the warm partially ionized component of the interstellar medium, given 
the relatively large flux of UV photons produced by radiative shock waves.
As we pointed out, the photoinizing flux produced by SNR's will be
particularly important in systems lacking massive stars.

{\bf Acknowledgments}
The excellent support received from C. Blondel, P. Mulet and the technical 
staff at San Pedro M\'artir observatory is gratefully acknowledged.
We thank the anonymous referee for the comments and suggestions
that lead to great improvements on this paper, and in particular for 
pointing out the effect of contamination from the other order to
the standard star continuum.


\begin{center} References \end{center}
\begin{description}
\item Ballet, J., Caplan, J., Rothenflug, R., Dubreuil, D. \& Soutoul, A.
1989, \aa 211 217
\item Ballet, J. \& Rothenflug, R. 1989, \aa 218 277
\item Binette, L., Dopita, M.A. \& Tuohy, I.R. 1985, \apj 297 476
\item Bochkarev, N.G. \& Sitnik, T.G. 1985, \apss 108 237
\item Brunish, W.M. \& Truran, J.W. 1982, \apjsupp 49 447
\item Chevalier, R.A. 1977, \annrev 15 175
\item Chevalier, R.A. 1990, in Supernovae, A\&A Library, ed. A.G. Petschek 
(Springer-Verlag, New York), 91
\item Cox, D.P. 1972, \apj 178 143
\item Cox, D.P. $\&$ Raymond, J.C. 1985, \apj 298 651
\item Cruz-Gonzalez, C., Recillas-Cruz, E., Costero, R., Peimbert, M. \&
Torres-Peimbert, S. 1974, \revmex 1 211
\item Daltabuit, E. \& Cox, D.P. 1972, \apj 173 L173
\item Decourchelle, A., Sauvageot, J.L., Ballet, J. \& Aschenbach, B. 1997, \aa 326
811
\item DeNoyer, L.K. 1975, \apj 196 479
\item Dopita, M.A. 1977, \apjsupp 33 437
\item Dopita, M.A. 1995, in The analysyis of emission lines, STScI Symp. Series
8, ed. R.E. Williams \& M. Livio (Cambridge, New York), 65
\item Dubreuil, D., Sauvageot, J.L., Blondel, C., Dhenain, G., Mestreau, P. \&
Mullet, P. 1995, Exp. Astron. 6, 257
\item Fransson, C., Cassatella, A., Gilmozzi, R., Kirshner, R.P., Panagia, N., 
Sonneborn, G. $\&$ Wamsteker, W. 1989, \apj 336 429
\item Garmany, C.D., Conti, P.S., Chiosi, C. 1982, \apj 263 777
\item Georgelin, Y.M., Lortet-Zuckerman, M.C. $\&$ Monnet, G. 1975, \aa 42 273
\item Graham, J.R., Wright, G.S., Hester, J.J. $\&$ Longmore, A.J. 1991, \aj 101 175
\item Green, D. A. 1988, \apss 148 3 
\item Hester, J.J., Raymond, J.C. \& Blair, W.P. 1994 \apj 1994, \apj 420 721
\item Kafatos, M., Sofia, S., Bruhweiler, F. $\&$ Gull, T. 1980, \apj 242 294
\item Ku, W.H.M., Kahn, S.M., Pisarski, R. \& Long, K.S. 1984, \apj 278 615
\item Levenson, N.A., Graham, J.R., Aschenbach, W.P., Blair, W.P.,
Brinkmann, W., Busser, J.U., Egger, R., Fesen, R.A., Hester, J.J.,
Kahn, S.M., Klein, R.M., McKee, C.F., Petre, R., Pisarski, R.,
Raymond, J.C. \& Snowden, S.L. 1997, \apj 484 304
\item McKee, C.F., Van Buren, D. $\&$ Lazareff, B. 1984, \apj 278 L115
\item Raymond, J.C. 1979, \apjsupp 35 419
\item Reynolds, R.J. 1983, \apj 268 698
\item Reynolds, R.J. 1985, \apj 298 L27
\item Reynolds, R.J. \& Tufte, S.L. 1995, \apj 439 L17
\item Sauvageot, J.L. \& Decourchelle, A. 1995, \aa 296 201
\item Sauvageot, J.L., Ballet, J., Dubreuil, D., Rothenflug, R., Soutoul, A.
\& Caplan, J. 1990, \aa 232 203
\item Scoville, N.Z., Irvine, W.M., Wannier, P.G. \& Predmore, C.R. 1977,
\apj 216 320
\item Sedov, L.I. 1959, Similarity and dimensional methods in mechanics, 
Academic Press, New York.
\item Shigeyama, T., Nomoto, K. $\&$ Hashimoto, M. 1988, \aa 196 141
\item Shull, J.M. \& McKee, C.F. 1979, \apj 227 131 
\item Stasinska, G. 1978 \aasupp 32 429
\item Sutherland, R.S. \& Dopita, M.A. 1993, \apjsupp 88 253
\item Sutherland, R.S., Bicknell, G.V. \& Dopita, M.A. 1993, \apj 414 510
\item Woosley, S.E. 1988, \apj 324 466
\end{description}

\newpage 
{\noindent {\bf Figure Captions}}
\begin{description}

\item Figure 1. [OII] lines: (a) sky, (b) region NE14.
\item Figure 2. [OIII] lines: (a) sky, (b) region NE18.
\item Figure 3. Observational pointings: (a) NE, (b) E and (c) SW. 
North is up, east is to the left.
\item Figure 4. [OII] and [OIII] NE and E traces.

\end{description}

\vbox{
\halign {\strut ~#~ \hfil & \hfil ~#~ \hfil & \hfil ~#~ \hfil & \hfil ~#~ 
\hfil & \hfil ~#~ \hfil & \hfil ~#~ \hfil & \hfil ~#~ \hfil & \hfil # \cr
\multispan 8 \strut {\bf Table 1. Northeastern trace~~~~~~~~~~~~~~~~~~~~~~~~~~~~
~~~~~~~~~~~~~~~~~~~~~~~~~~~~~~~~~~~~~~} \hfil  \cr
\raya
~ \col RA  (1950)  DEC \col $\delta(")$ \col 1+ $\delta/\theta$ 
\col I([OII]3729) \col I([OIII]5007) \col 3729/5007 \cr
\raya
NE0 \col  20 53 55.8 +31 35 05  \col  -884  \col 0.85  \col  ~
\col 30.1 \menmas  6.1  6.1 \col ~ \cr  

NE1 \col  20 54 05.0 +31 37 35 \col -663 \col 0.89  \col 79.0  \menmas 0.0  0.0    \col 67.8 \menmas 9.3 9.3 \col  1.17 \menmas 0.14 0.18 \cr

NE2 \col  20 54 20.0 +31 39 20 \col -390 \col 0.93 \col  2.91 \menmas 0.34 0.23   \col 2.89 \menmas 0.60 0.21 \col 1.01 \menmas 0.18 0.37 \cr 

NE3 \col  20 54 19.0 +31 41 05 \col -325 \col 0.94  \col 1.71 \menmas 0.23 0.19  \col 1.83 \menmas 0.31 0.36  \col 0.93 \menmas 0.26 0.32 \cr 

NE4 \col  20 54 26.0 +31 42 50 \col -162 \col 0.97  \col 1.13 \menmas 0.22 0.16  \col 1.38 \menmas 0.33 0.33 \col  0.82 \menmas 0.30 0.41 \cr  
  
NE5 \col  20 54 33.0 +31 44 35  \col   0 \col 1 \col 1.02 \menmas 0.23 0.18 
\col 0.85 \menmas 0.24 0.24  \col 1.20 \menmas 0.47 0.76 \cr
  
NE6 \col  20 54 40.0 +31 46 20  \col 162 \col 1.03 \col 0.90 \menmas 0.10 0.20    \col 0.99 \menmas 0.35 0.48 \col 0.91 \menmas 0.37 0.82  \cr
      
NE7  \col 20 54 47.0 +31 48 05  \col 325 \col 1.06  \col 0.80 \menmas 0.21 0.15 
\col 0.96 \menmas 0.25 0.24  \col 0.83 \menmas 0.34 0.50  \cr
  
NE8  \col 20 54 54.0 +31 49 50 \col  487 \col 1.08  \col 0.67 \menmas 0.19 0.13  \col 0.71 \menmas 0.19 0.15  \col 0.94 \menmas 0.38 0.58  \cr

NE9 \col 20 55 01.0 +31 51 35 \col  650 \col 1.11 \col  0.78 \menmas 0.23 0.29 
\col 0.73 \menmas 0.21 0.26 \col  1.07 \menmas 0.42 1.02 \cr
 
NE12 \col 20 55 22.0 +31 54 50  \col 1137 \col 1.19  \col 0.65 \menmas 0.21 0.10
\col ~   \col ~ \cr

NE14 \col 20 55 31.0 +32 00 00 \col 1485 \col 1.25  \col 0.70 \menmas 0.21 0.10 
\col ~  \col ~ \cr   

NE15 \col 20 55 43.0 +32 02 05 \col 1702 \col 1.29  \col 0.77 \menmas 0.25 0.20  \col 0.65 \menmas 0.16 0.19 \col  1.18 \menmas 0.56 0.88 \cr
  
NE18 \col 20 56 04.0 +32 07 20 \col 2189 \col 1.37  \col 0.83 \menmas 0.18 0.13    \col 0.85 \menmas 0.31 0.31  \col 0.98 \menmas 0.42 0.81   \cr  

NE22 \col 20 56 23.0 +32 17 02 \col 2754 \col 1.47 \col 0.65 \menmas 0.15 0.20
\col 0.95 \menmas 0.24 0.30 \col 0.69 \menmas 0.37 0.33  \cr
\raya
}}
\begin{description}
\item Intensity in 10$^{-7} erg~cm^{-2}~s^{-1}~sr^{-1}$ 
\end{description}

\vbox{
\halign {\strut ~#~ \hfil & \hfil ~#~ \hfil & \hfil ~#~ \hfil & \hfil ~#~ 
\hfil & \hfil ~#~ \hfil & \hfil ~#~ \hfil & \hfil ~#~ \hfil & \hfil # \cr
\multispan 8 \strut {\bf Table 2. Eastern trace~~~~~~~~~~~~~~~~~~~~~~~~~~~~~~~~~
~~~~~~~~~~~~~~~~~~~~~~~~~~~~~~~~~~~~~~} \hfil  \cr
\raya
~ \col RA  (1950)  DEC \col $\delta(")$ \col 1+ $\delta/\theta$ 
\col I([OII]3729) \col I([OIII]5007) \col 3729/5007 \cr
\raya

E0 \col 20 55 13.8  +30 54 30 \col -705 \col 0.88  
\col 45.20 \menmas 1.70 1.30 \col 198.0 \menmas 15.8 15.8 \col 0.23 
\menmas 0.03 0.02 \cr     

E1 \col 20 55 24.0  +30 54 30 \col -531 \col 0.91 
\col 18.20 \menmas 1.10 0.40  \col 111.3 \menmas 11.8 11.8 \col 0.16 
\menmas 0.02 0.02 \cr         

E2 \col 20 55 34.0  +30 54 30 \col -354  \col 0.94  
\col 2.98 \menmas 0.27 0.36  \col 4.86 \menmas 0.25 0.24 \col 0.62
\menmas 0.09 0.10 \cr           

E3  \col 20 55 44.0  +30 54 30 \col -177  \col 0.97  
\col 1.45 \menmas 0.20 0.03  \col 2.04 \menmas 0.70 0.34 \col 0.71 
\menmas 0.18 0.39 \cr         

E4 \col 20 55 54.0  +30 54 30 \col    0  \col 1.     
\col 1.54 \menmas 0.35 0.10  \col 1.43 \menmas 0.25 0.29 \col 1.08 
\menmas 0.38 0.31 \cr        

E5  \col 20 56 04.0  +30 54 30 \col  177  \col 1.03   
\col 1.56 \menmas 0.47 0.14  \col 1.61 \menmas 0.35  0.43 \col 0.96 
\menmas 0.42 0.38 \cr

E6  \col 20 56 14.0  +30 54 30 \col  354  \col 1.06   
\col 1.15 \menmas 0.29 0.22  \col 2.14 \menmas 0.24  0.34  \col 0.54 
\menmas 0.19 0.18 \cr        

E8  \col 20 56 34.0  +30 54 30 \col  708   \col 1.12   
\col 0.91 \menmas 0.19 0.16   \col 1.09  \menmas 0.33  0.26 \col 0.84 
\menmas 0.30 0.56 \cr                 

E10   \col 20 56 54.0  +30 54 30 \col 1062 \col  1.18   
\col 0.78 \menmas 0.24 0.15  \col 1.04 \menmas 0.39  0.18   \col 0.75 
\menmas 0.30 0.68 \cr       

E14  \col 20 57 34.0  +30 54 30 \col 1770 \col  1.30   
\col 0.77 \menmas  0.14 0.11  \col 0.93 \menmas 0.21  0.23  \col 0.83 
\menmas 0.29 0.40 \cr
         
E18  \col 20 58 14.0  +30 54 30 \col 2478 \col  1.42   
\col 0.81 \menmas 0.19 0.10  \col 0.45 \menmas 0.16 0.18   \col 1.80 
\menmas 0.81 1.37 \cr           

E26 \col 20 59 34.0  +30 54 30 \col 3894 \col  1.67   
\col 0.61 \menmas 0.08 0.16  \col 0.46 \menmas 0.15 0.20 \col 1.31 
\menmas 0.29 1.15 \cr  
\raya
}}
\begin{description}
\item Intensity in 10$^{-7} erg~cm^{-2}~s^{-1}~sr^{-1}$ 
\end{description}

\vbox{
\halign {\strut ~#~ \hfil & \hfil ~#~ \hfil & \hfil ~#~ \hfil & \hfil ~#~ 
\hfil & \hfil ~#~ \hfil & \hfil ~#~ \hfil & \hfil ~#~ \hfil & \hfil # \cr
\multispan 8 \strut {\bf Table 3. Southwestern trace~~~~~~~~~~~~~~~~~~~~~~~~~~~~~~~~~
~~~~~~~~~~~~~~~~~~~~~~~~~~~~~~~~~~} \hfil  \cr
\raya
~ \col RA  (1950)  DEC \col $\delta(")$ \col 1+ $\delta/\theta$ 
\col I([OII]3729) \col I([OIII]5007) \col 3729/5007 \cr
\raya

SW0 \col 20 44 55.3  +30 00 36 \col -631 \col .89 
\col 7.02 \menmas 0.52 0.48 \col 15.1 \menmas 0.8 0.9 \col 0.46 
\menmas 0.06 0.06 \cr     

SW1 \col 20 44 35.3  +29 57 36 \col -314 \col 0.95
\col   \col 48.5  \col  \cr         

SW1.5 \col 20 44 25.3  +29 56 08 \col -158  \col 0.97 
\col 1.71 \menmas 0.20 0.16  \col 1.66 \menmas 0.64 0.35 \col 1.03 
\menmas 0.28 0.79 \cr           

SW2  \col 20 44 15.3  +29 54 36 \col 0 \col 1  
\col 1.14 \menmas 0.17 0.22  \col 0.78 \menmas 0.30 0.54 \col 1.47 
\menmas 0.74 1.38 \cr         
      
SW4  \col 20 43 35.3  +29 48 36 \col    633  \col 1.11     
\col 0.85 \menmas 0.18 0.23  \col 0.89 \menmas 0.34 0.16 \col 0.96 
0.32 1.00 \cr        
\raya
}}
\begin{description}
\item Intensity in 10$^{-7} erg~cm^{-2}~s^{-1}~sr^{-1}$ 
\end{description}

\vbox{
\halign {\strut ~#~ \hfil & \hfil ~#~ \hfil & \hfil ~#~ \hfil & \hfil 
                ~#~ \hfil & \hfil ~#~ \hfil & \hfil ~#~  \hfil & \hfil 
               ~#~ \hfil & \hfil ~#~ \hfil & \hfil ~#~ \hfil & \hfil # \cr
\multispan 9 \strut {\bf Table 4. Photoionizing shock}
~~~~~~~~~~~~~~~~~~~~~~~~~~~~~~~~~~~~~~~~~~~~~~~~~ \hfil  \cr
\raya
      \col $E_{50}=1$, \col $N_0=1$
      \col $E_{50}=3$, \col $N_0=0.2$
      \col $E_{50}=1$, \col $N_0=0.2$ \cr
$V_7$ \col R$_{pc}$         \col R$_i$ (pc) 
      \col R$_{pc}$         \col R$_i$ (pc) 
      \col R$_{pc}$         \col R$_i$ (pc) \cr
\raya
1.0   \col 19 \col 27 
      \col 48 \col 79
      \col 33 \col 61 \cr
1.5   \col 15 \col 36 
      \col 37 \col 112
      \col 25 \col 87 \cr
2.0   \col 12 \col 44 
      \col 30 \col 136 
      \col 21 \col 107 \cr
2.5   \col 11 \col 50
      \col 26 \col 157
      \col 18 \col 123 \cr     
\raya
}}

\begin{figure}
\psfig{file=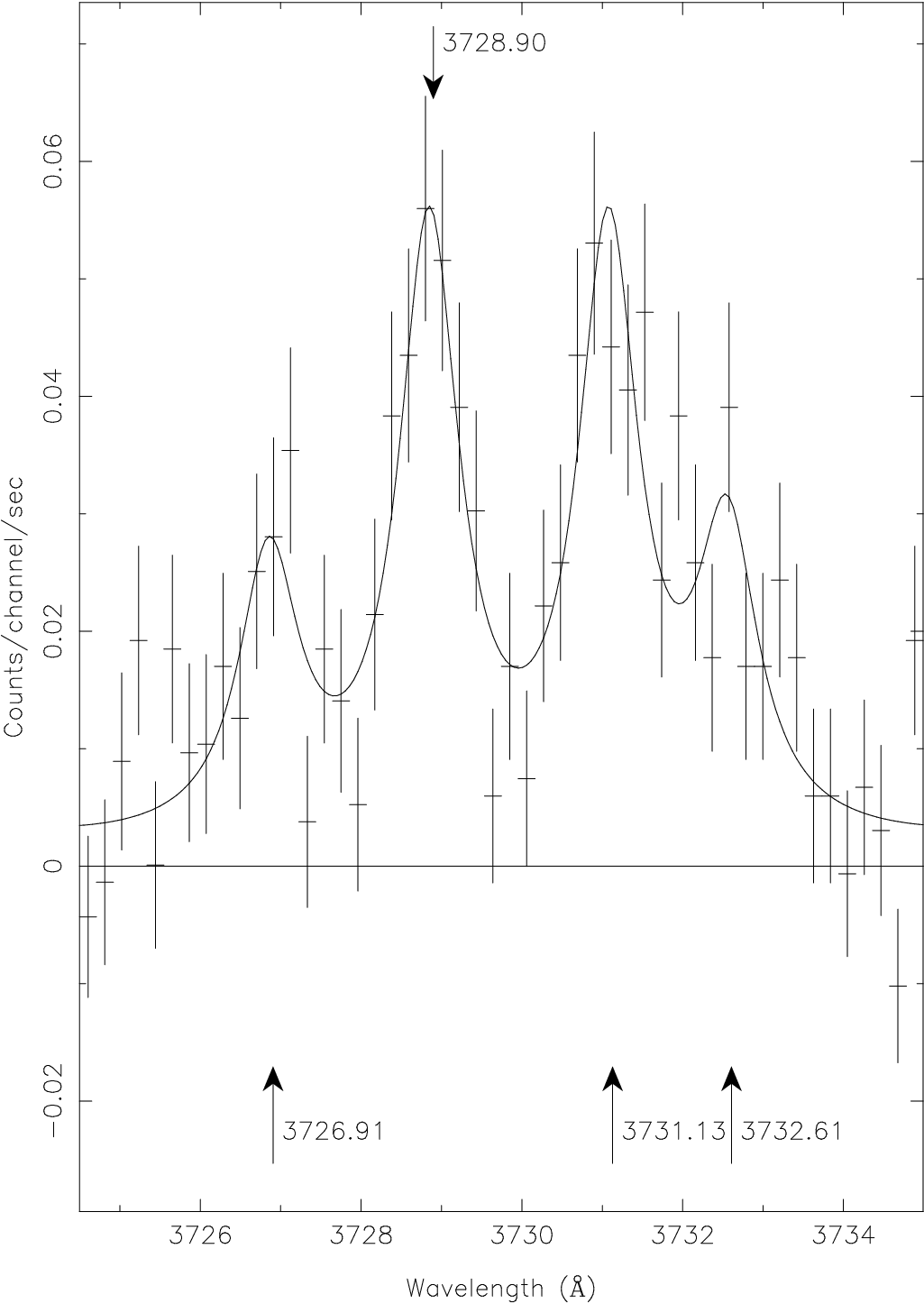,height=11cm,width=17cm}
Figure 1a.\\
\psfig{file=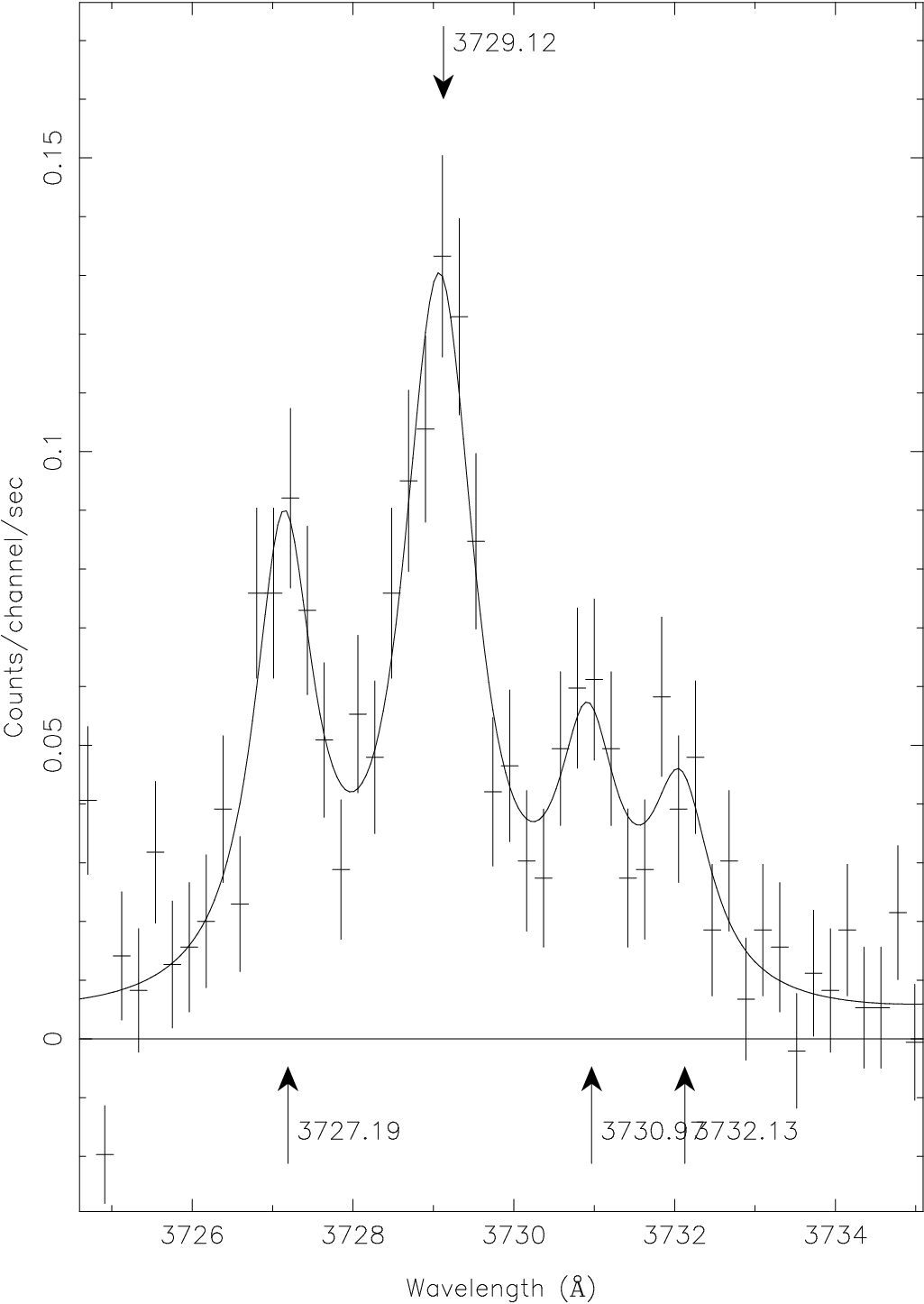,height=11cm,width=17cm}
Figure 1b.\\
\end{figure}

\begin{figure}
\psfig{file=fig2a.ps,height=11cm,width=17cm}
Figure 2a.\\
\psfig{file=fig2b.ps,height=11cm,width=17cm}
Figure 2b.\\
\end{figure}

\begin{figure}
\psfig{file=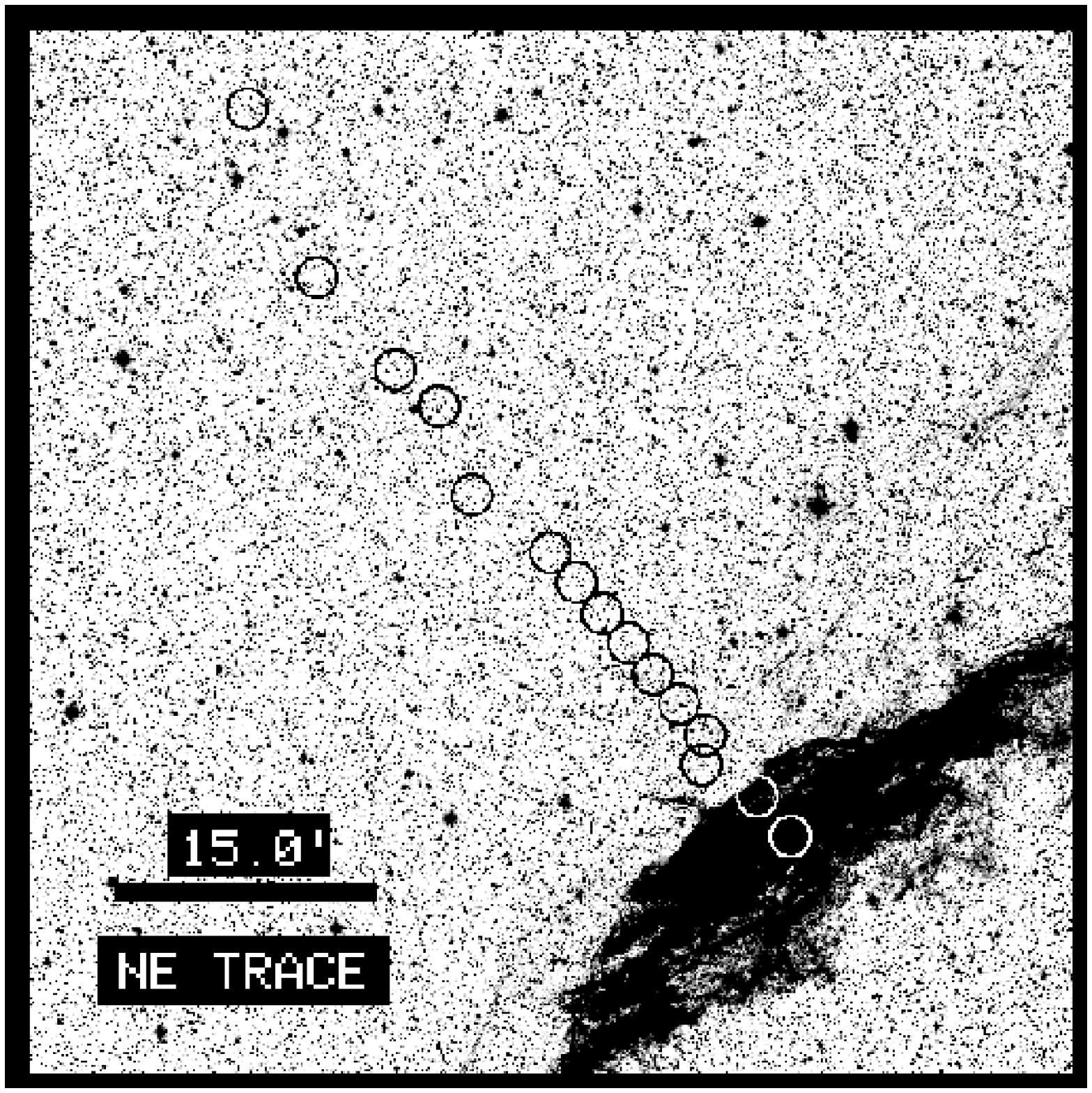}
Figure 3a.\\
\end{figure}
\begin{figure}
\psfig{file=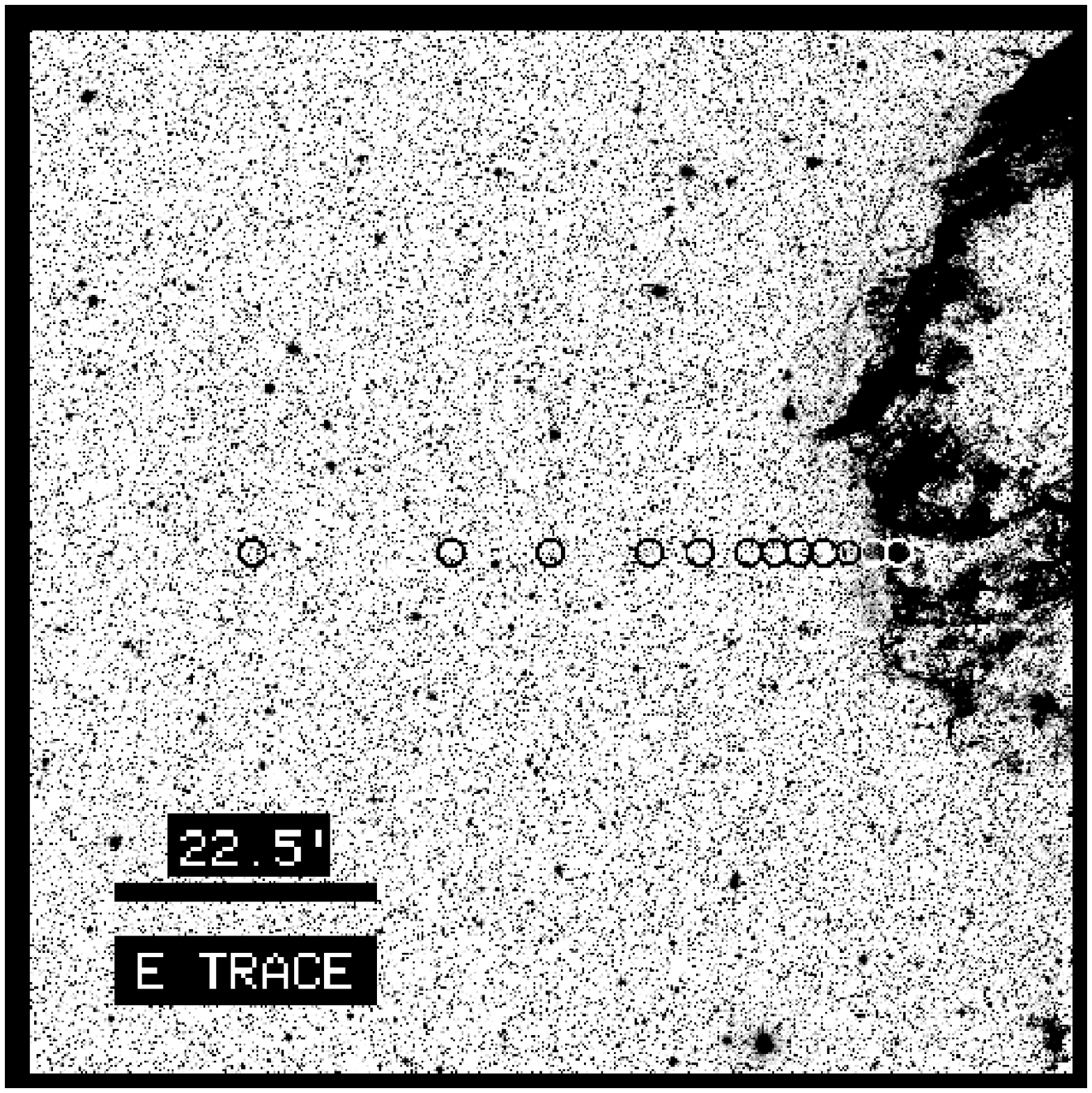}
Figure 3b.\\
\end{figure}
\begin{figure}
\psfig{file=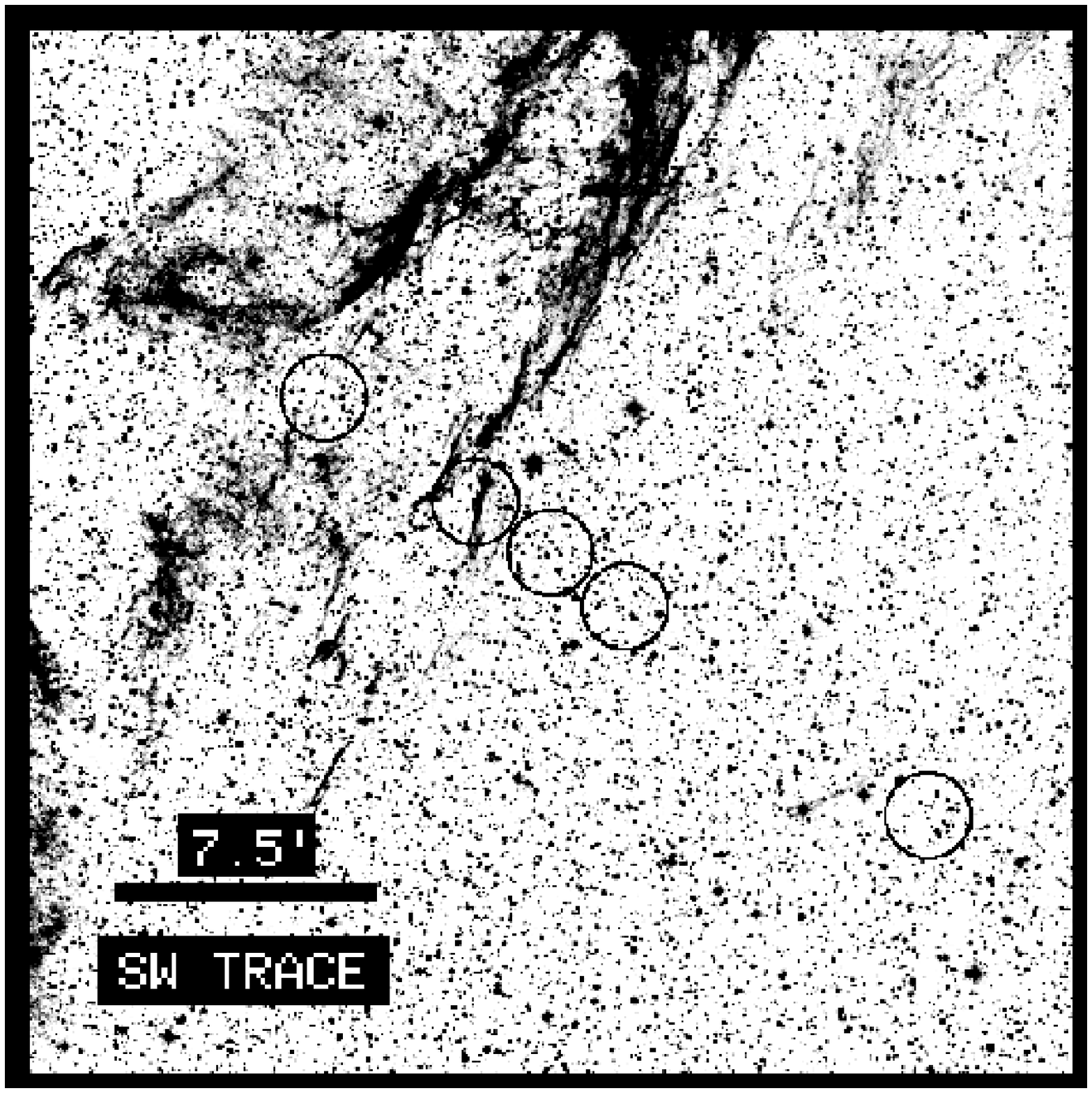}
Figure 3c.\\
\end{figure}

\begin{figure}
\plotone{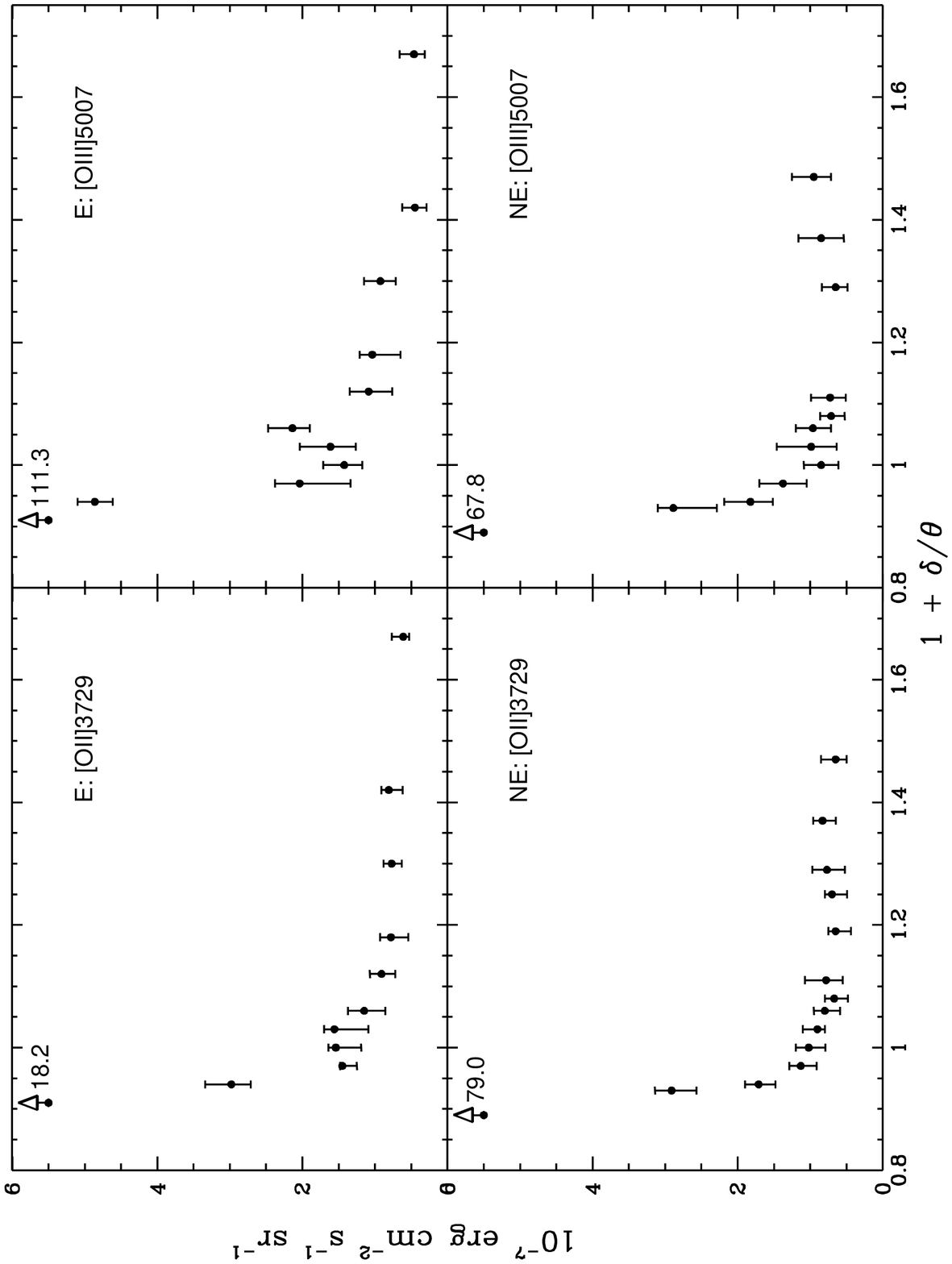}
Figure 4.\\
\end{figure}

\end{document}